\documentclass[10pt,twocolumn,letterpaper]{article}

\usepackage[pagenumbers]{iccv} 

\usepackage{multirow}
\usepackage{tabularx}

\usepackage{listings}
\lstdefinestyle{mystyle}{
    basicstyle=\ttfamily\footnotesize,
    backgroundcolor=\color{gray!10},
    breaklines=true,  
    frame=single,     
    captionpos=t      
}
\definecolor{iccvblue}{rgb}{0.21,0.49,0.74}
\usepackage[pagebackref,breaklinks,colorlinks,allcolors=iccvblue]{hyperref}
\usepackage{censor}
\usepackage{graphicx}
\usepackage{float}
\usepackage{tikz}
\usepackage{caption}
\usetikzlibrary{arrows.meta,
                calc, chains,
                quotes,
                positioning,
                shapes.geometric}
\def\paperID{4} 
\def\confName{ICCV AISTORY Workshop}
\def\confYear{2025}

\title{From Sound to Sight: Towards AI-authored Music Videos}

\author{Leo Vitasovic\\
IT University of Copenhagen\\
{\tt\small leov@itu.dk}
\and
Stella Graßhof\\
IT University of Copenhagen\\
{\tt\small stgr@itu.dk}
\and
Agnes Mercedes Kloft\\
Aalto University\\
{\tt\small agnes.kloft@aalto.fi}
\and
Ville V. Lehtola\\
University of Twente\\
{\tt\small v.v.lehtola@utwente.nl}
\and
Martin Cunneen\\
University of Limerick\\
{\tt\small martin.cunneen@ul.ie}
\and
Justyna Starostka\\
IT University of Copenhagen\\
{\tt\small juss@itu.dk}
\and
Glenn McGarry\\
University of Nottingham\\
{\tt\small glenn.mcgarry@nottingham.ac.uk}
\and
Kun Li\\
University of Twente\\
{\tt\small k.li@utwente.nl}
\and
Sami S. Brandt\\
IT University of Copenhagen\\
{\tt\small sambr@itu.dk}
}

\begin{document}

\twocolumn[{%
\renewcommand\twocolumn[1][]{#1}%
\maketitle
\begin{center}
    \centering
   \includegraphics[width=1\textwidth, trim=40 200 50 0, clip]{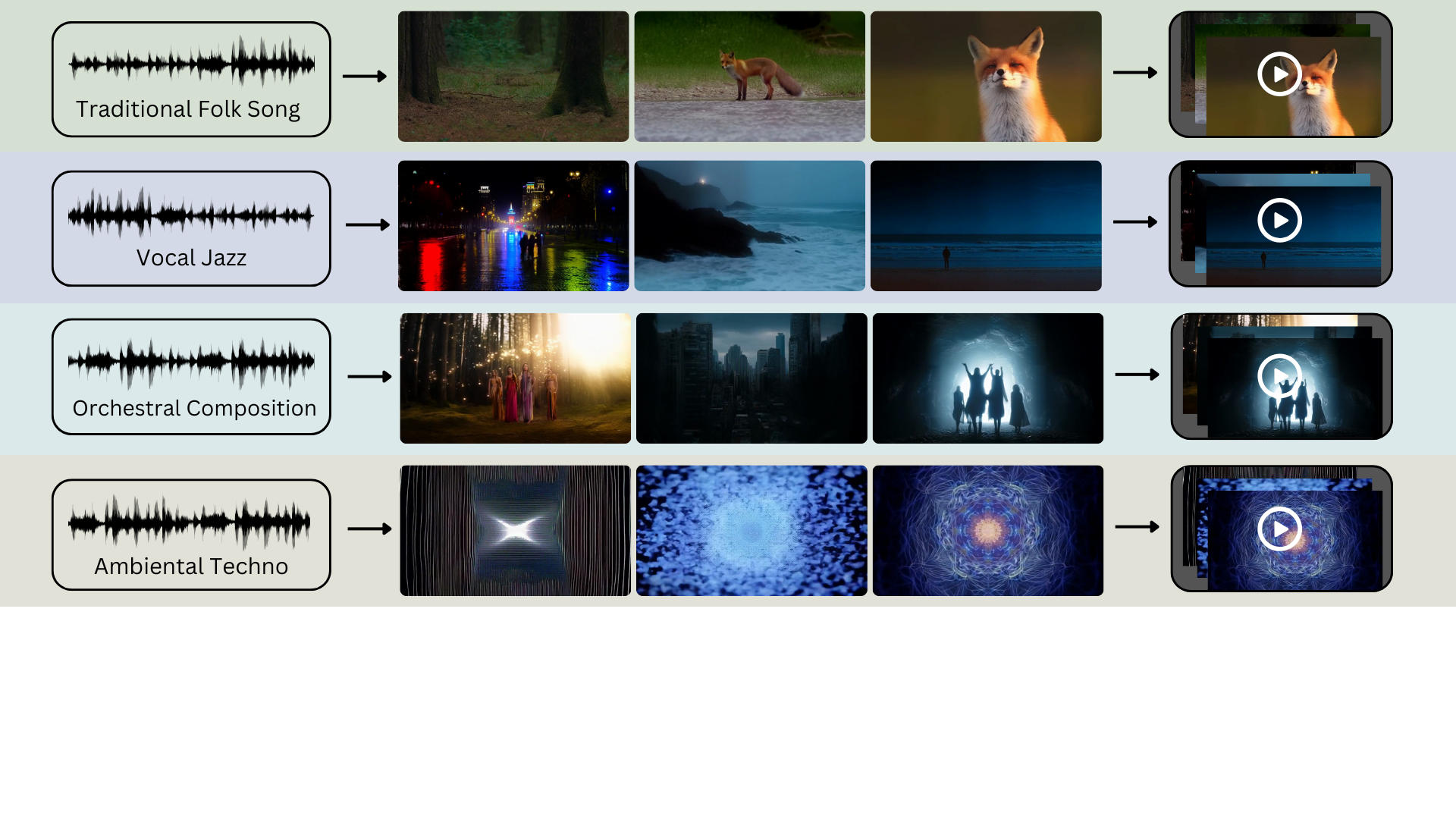}
    \captionof{figure}{Overview of the method proposed in this paper. Given a selected song, the pipeline generates video clips, which are then merged into a music video. The figure includes screenshots from clips generated given the specified music. The examples feature the following songs (from top to bottom): Dónal O'Connor and Muireann Nic Amhlaoibh: "Fairy Jig", Frank Sinatra: "Strangers in the Night", "The March of the Volunteers" (Chinese national anthem) composed by Niè Ěr, and Aphex Twin: "Xtal", which can all be found on our \href{https://github.com/goodPointP/Results-For-Music-Visualization-Generation-Pipeline}{Github}.}
    \label{fig:teaser}
\end{center}
}]



\begin{abstract}
Conventional music visualisation systems rely on handcrafted ad hoc transformations of shapes and colours that offer only limited expressiveness. We propose two novel pipelines for automatically generating music videos from any user-specified, vocal or instrumental song using off-the-shelf deep learning models. Inspired by the manual workflows of music video producers, we experiment on how well latent feature-based techniques can analyse audio to detect musical qualities, such as emotional cues and instrumental patterns, and distil them into textual scene descriptions using a language model. Next, we employ a generative model to produce the corresponding video clips. To assess the generated videos, we identify several critical aspects and design and conduct a preliminary user evaluation that demonstrates storytelling potential, visual coherency and emotional alignment with the music. Our findings underscore the potential of latent feature techniques and deep generative models to expand music visualisation beyond traditional approaches.
\end{abstract}    
\section{Introduction}
This paper explores the processes of creating meaningful visuals that support storytelling to accompany a piece of music using AI-generated videos. Storytelling has been an important characteristic of social development; oral traditions were once a key source of information communication by contextualising valuable information in a medium of a story \cite{bortolussi_review_2012,rubin_memory_1995}. For some people, the ability of the story to conjure images and visuals provided a stimulating means of learning and information acquisition \cite{graesser_constructing_1994}. The underlying idea of this paper is based on the natural tendency of people to associate music with other modalities, such as sight or touch \cite{iosifyan_cross-modal_2022,bulusu_crossmodal_2024}. Such innate cross-modal perception may have influenced early attempts to visually augment music, such as George H. Thomas’s creation of a series of images that accompanied the live performance of The Little Lost Child in 1892, using a stereopticon \cite{goldberg_witmark_ragtime_1939}. At the same time, on the other side of the world, Jean Sibelius (1865–1957) experienced synaesthesia — a condition in which he perceived sounds and musical keys as vivid colours \cite{pearce2007synaesthesia}. This blending of the senses enabled him not only to hear the natural world but also to see it in music, translating the landscapes of his homeland into compositions that evoke striking visual imagery in the listener’s mind.

From the days of the stereopticon to the 20th and 21st centuries, the evolution of audiovisual technology, from the film camera and broadcast television to personal computers and the internet, has continuously reshaped how music is experienced. These innovations paved the way for music videos, transforming them into a widely recognised art form that enhances the listener's perception of music through visual storytelling \cite{kniaz-hunek_revolution_2023,dasovich-wilson_exploring_2022}. What began as individual sensory experiences, such as Sibelius’s synaesthetic perceptions or Thomas’s early visual accompaniments, has evolved into a global multimedia phenomenon, demonstrating the growing synergy between sound and vision in artistic expression.

AI-generated art has evolved from simple automated creations to complex works spanning across multiple artistic domains \cite{audry_art_2021, cetinic_understanding_2021, sivertsen_machine_2024}. This transformation has been largely driven by advancements in deep learning, particularly in generative models such as Generative Adversarial Networks (GANs) \cite{goodfellow_generative_2020} and diffusion models \cite{boutin_diffusion_2023}. Today, AI-generated works have been exhibited in galleries and have gained public attention in art auctions. The development and attention challenge traditional notions of artistic authorship and creativity \cite{epstein_who_2020}. In the field of music visualisation, AI follows similar trends, automating what was once a handcrafted process. Traditional methods relied on manual design to translate music into visuals \cite{lima_survey_2022}. AI-powered tools, however, introduce new solutions by automating the process, generating synchronised visuals from audio input. For instance, text-to-image diffusion models can translate musical themes and lyrics into visual sequences, but artists must manually specify the input text for the models, e.g. \cite{agarwal_secure_2025}, as fully automated solutions remain relatively sparse \cite{tunebat_llc_specterr_2025,kaiber_corp_kaiber_2025}.

This shift raises fundamental questions about the autonomy of AI in artistic creation. Some scholars argue that AI systems function only as tools manipulated by human artists \cite{elgammal_can_2017, hertzmann_can_2018, hertzmann_computers_2020}, while others claim that AI possesses a degree of creative autonomy, influencing the artistic process beyond mere execution \cite{boden_what_2009}. Our research broadly aims to explore the potential for fully automating the art creation process, contributing to this ongoing debate about the agency of human creators versus AI models. Moreover, audience perception of AI-generated art remains an open discussion, as different studies show that human-created artworks are generally rated higher in expressiveness compared to AI-generated ones \cite{hong_artificial_2019, horton_jr_bias_2023}. Ethical concerns, particularly regarding training data and originality, further complicate debates on creative ownership of AI-generated artworks \cite{cetinic_understanding_2021}.

Art is created for human experience \cite{cetinic_understanding_2021}, and its creation has traditionally been an exclusively human domain \cite{dewey_art_2005}. The introduction of generative AI into this area is not merely a technological shift but an ethical one, challenging the value of human skill and creative labour. Indeed, new advances in generative AI blur the lines between human-made, human-curated, and human-inspired art \cite{mazzone_art_2019,epstein_art_2023}. This ambiguity creates ethical and legal challenges regarding authorship and copyright, as it becomes difficult to disentangle the contributions of the user, the AI model, and the creators of the original training data \cite{kevins_artificial_2021}. Consequently, the ongoing integration of AI into the arts necessitates a deeper ethical framework to address issues of labour devaluation, creative authenticity, and the ownership of computationally generated culture.

One of the most technically developed, yet ethically sensitive, applications of audio-to-video generation is speech-to-lip synchronisation, often used to create photorealistic talking heads \cite{kumar_robust_2020}. In contrast, our work focuses on a less sensitive and less explored task: generating video directly from music to support creative storytelling. 
While text-to-video generation has advanced significantly with the rise of latent diffusion models and transformer-based architectures \cite{arkhipkin_improveyourvideos_2025}, audio-to-video synthesis from music remains an open research challenge. 
Early forms of music visualisation, such as those found in Windows Media Player \cite{microsoft_corporation_windows_2025}, rely on signal processing techniques like Fourier transforms and spectrograms to create reactive animations. 
More recent tools, such as Specterr \cite{tunebat_llc_specterr_2025} or Kaiber \cite{kaiber_corp_kaiber_2025}, offer stylised audio-reactive visuals via templates and simple heuristics. However, these systems lack semantic understanding, narrative structure, or temporal visual consistency. Their methodologies are proprietary and scientifically undocumented. 
Similarly, several commercial tools use AI for audio-to-video generation, including Revid \cite{tmaker_revidai_2025}, NeuralFrames \cite{neuralframes_ai_2025}, and EasyVid \cite{easyvid_easyvid_2024}. 
However, none provide technical details, and many rely heavily on lyrics or the additional user-input via text-prompts for content and style guidance, hence limiting their generalizability to audio without lyrics. 
A related studied domain is body motion synchronisation with music, such as generating performance videos of people playing instruments or dancing. For instance, Zhu et al. \cite{zhu_lets_2021} present a pipeline for synthesising videos of instrumental performances from raw audio. 
Similarly, Ren et al. introduce a method capable of aligning a generated dance sequence with the beat and rhythm of a song \cite{ren_self-supervised_2020}.  Those approaches aim to synchronise audio and generated human motion, smoothly capturing rhythmic and stylistic features. However, they are typically restricted to controlled human poses and fixed camera settings, hence hindering broader scene synthesis or narrative depth. 

One recent solution towards music-to-video storytelling has been proposed by Agarwal et al. \cite{agarwal_secure_2025}, who offer a pipeline that generates music videos using lyrics, estimated emotional tone, and user style preferences. 
Their method relies on Whisper \cite{radford_robust_2022} to extract lyrics, followed by emotion estimation and LLM-based text refinement \cite{openai_hello_2025}. The resulting text is used as input for Stable Diffusion \cite{rombach_high-resolution_2022} to generate images, which are interpolated into a cohesive video. However, the method is limited to music with lyrics, and its visual coherence is heavily dependent on textual accuracy. 

In contrast, our work is designed to generalise to any musical input, including instrumental and non-verbal audio, enabling music-first video generation without reliance on lyrics. 
Inspired by the human workflow used by music video creators, our approach to \emph{computational synaesthesia} aims to build on this foundation to provide appealing audio and music visualisations, as shown in Figure \ref{fig:teaser}. 

The pipelines we propose present a novel approach to achieve computational synaesthesia in a technically sophisticated way that is also guided by (a) AI governance and ethics, and (b) artistic considerations to support greater value alignment in the model. In this way, the model design is informed by an ethics-by-design framework that combines the considerations from technical, artistic and governance points of view. The research takes a human-centric approach to computational synaesthesia by appealing to five key design features; (i) developing the pipeline on a human workflow, (ii) using natural language as the key medium of processing, (iii) informing the pipeline by the defining component of human synaesthesia as sensory transformation, (iv) informing the pipeline with Artist values and (v) informing the approach with AI governance, risk and ethical assessment. Note that one of our key design choices involves the use of text as a medium to keep the interpretability of the AI methods high, including being able to comply with the artist's values and AI ethical standards.

The contributions of the paper are as follows.
\begin{itemize}
    \item \textbf{Instrumental Music Visualisation} Unlike lyric-based methods, this approach only relies on instrumental cues to drive visuals. This enables video generation for instrumental and lyrics-featuring music alike.
    \item \textbf{Latent Feature Techniques for Audio–Text Alignment} We explore the potential of using contrastive language-audio pre-training (CLAP) and large audio language models (LALM) to extract zero-shot, high-level musical attributes to represent an audio piece and write a story inspired by the music.
    \item \textbf{LLM-Based Scene Scripting} We use a large language model to translate CLAP-derived descriptors into concise, narrative-like scene prompts, guiding text-to-video generation.
    \item \textbf{Degree of AI Agency in Art Creation} We pose questions about the degree of freedom with which an automated pipeline can generate art independently or with minimal guidance, and the cultural and artistic qualities and challenges the results imply.
    \item \textbf{AI Music Video Evaluation} We design and conduct a user survey to evaluate the quality of storytelling and visual content of the generated samples. We complement this survey by holding a more in-depth interview on the AI video generation capabilities of the pipelines.
    \item \textbf{Code} is published on \href{https://github.com/goodPointP/Results-For-Music-Visualization-Generation-Pipeline}{Github}\footnote{https://github.com/goodPointP/Results-For-Music-Visualization-Generation-Pipeline}.
\end{itemize}
\section{Method}
We base our pipelines on existing models and systems described in this section.
\label{sec:method}
\subsection{Contrastive Language-Audio Pre-training}
\label{subsec:clap}

Contrastive language--audio pretraining (CLAP) \cite{elizalde_clap_2023} 
leverages contrastive learning to align audio signals and natural language descriptions in a joint embedding space. CLAP was trained on 128k audio-text pairs and evaluated on 16 downstream tasks spanning 8 different domains, demonstrating its versatility and robustness in modelling audio concepts. 
We leverage the foundation model in our pipeline for zero-shot audio analysis based on predefined class labels that we manually specify. 
Given an arbitrary musical input, vocal or instrumental, CLAP generates semantic labels which describe the audio’s characteristics, e.g. as `melodic piano', `upbeat tempo', or `sad and moody strings'. 
These high-level textual descriptors encapsulate the musical content without the need for large domain-specific datasets or extensive manual labelling, and therefore are particularly valuable for music visualisation tasks. 
They are later used by a large language model to construct scene descriptions and narrative elements for the output video. 


\subsection{Large Audio Language Models}
LALMs represent an emerging paradigm in multimodal artificial intelligence, designed to process and comprehend raw audio signals in conjunction with natural language \cite{ghosh_gama_2024}. Unlike conventional audio analysis techniques that rely on predefined feature extraction, LALMs are trained on extensive datasets of aligned audio and text, enabling them to develop a holistic understanding of complex auditory information, including musical structure, emotional valence, genre characteristics, and implicit narrative potential \cite{deshmukh_pengi_2023}. 
This integrated comprehension facilitates a semantically rich interpretation of audio, bridging the gap between acoustic phenomena and linguistic description.

In one of our two pipelines, we harness the advanced capabilities of LALMs to directly generate a coherent narrative concept or short story that thematically and emotionally resonates with a given input song. This method diverges from the CLAP-based approach by providing the raw audio track directly to the LALM, thereby circumventing the need for explicit, pre-extracted audio features. The LALM is prompted to synthesise a narrative concept that thematically and emotionally follows or would fit with the given song. This task evaluates the LALM's interpretive and generative capacities: its ability to infer abstract concepts such as mood, energy, and temporal progression from audio data, and subsequently translate these inferences into a structured, imaginative narrative suitable for a music video.

LALMs show potential to generate narratively coherent and emotionally aligned textual scripts with the given musical piece. The resulting narrative is then either directly translated into scene descriptions by the LALM itself or further processed by a reasoning Large Language Model to decompose the narrative into concrete, segment-aligned visual prompts for the subsequent text-to-video generation stage.

\subsection{Large Language Models}

Recent advances in large language models (LLMs) have made them adept at performing a wide range of tasks, from summarisation to complex reasoning and creative text generation \cite{liu_revisiting_2023,lee_reasoning_2025,yuan_wordcraft_2022}. In our workflow, we utilise LLMs specifically as a “video script-writing tool,” responsible for transforming CLAP-derived audio descriptors into coherent and contextually rich textual scene outlines.

We used \texttt{DeepSeek-R1-Distill-Llama-8B} by DeepSeek \cite{deepseek-ai_deepseek-r1_2025}, a reasoning-LLM \cite{wei_chain--thought_2023} that combines the efficiency of a distilled model \cite{xu_survey_2024} with the advanced reasoning capabilities typical of larger parameter models \cite{lu_are_2024}. We find that the said model can interpret nuanced audio concepts from CLAP’s output classes and use this knowledge to generate a detailed narrative structure that aligns with the mood and style of the music.

This reasoning step ensures that the generated video scenes are not merely random visual montages but are instead guided by a coherent storyline or thematic arc, reflecting the emotional landscape of the audio. The LLM-based approach makes it easy to iterate and refine prompts by adjusting textual descriptions or keywords, thus offering flexibility in shaping the final visual output.

\subsection{Text to Video Models}

The last stage of our pipeline involves converting the refined textual prompts into video clips using diffusion-based text-to-video models. Diffusion models have recently gained popularity for their ability to generate high-fidelity images and videos by iteratively denoising random noise toward a target distribution \cite{ho_cascaded_2021,ho_video_2022}. However, when extending diffusion techniques from images to videos, several challenges arise, such as limits in clip length and resolution, input prompt sensitivity and quality of human faces and overall image consistency. Obtaining realistic human faces remains particularly difficult; many models tend to produce distorted or unstable facial features, which can be distracting or diminish the overall video quality \cite{huang_vbench_2023}. Moreover, we find the outputs to be dependent on the concise wording of the input textual prompt. This quality poses a challenge for this method, as the prompts themselves are generated by another model, without human supervision. In our work, we utilise two models; \texttt{mochi-1} by GENMO \cite{genmo_team_mochi_2024} for the first pipeline, and \texttt{WAN 2.1} \cite{wan_wan_2025} for the second approach.
\begin{figure*}[ht]
\centering
\includegraphics[width=0.55\linewidth]{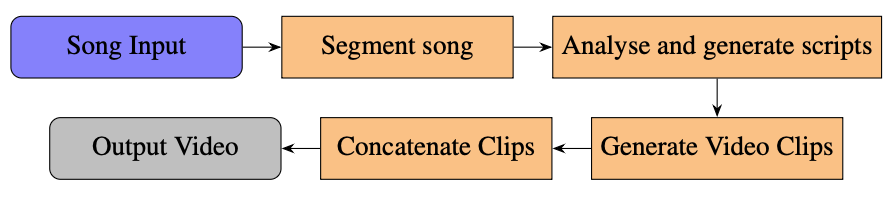}
\caption{Overview of the pipelines. The video for a selected song is generated through four main steps: (1) segmenting the song, (2) audio analysis and script generation, (3) generating video clips using text-to-video models from the script, and (4) assembling the final video.}
\label{fig:pipeline_overview}
\end{figure*}

\section{Two Audio to Video Pipelines}
We propose two pipelines that utilise tools presented in Sec. ~\ref{sec:method}. Both pipelines follow the steps described under Fig.~\ref{fig:pipeline_overview}, with slight variations regarding audio analysis and video script generation. They use off-the-shelf models and produce human-readable, interpretable intermediary steps.\footnote{The intermediary results, including the segmentation results, LLM prompts and the resulting scripts, are saved as individual text files.}

\subsection{Pipeline 1: CLAP-based approach}
The first pipeline utilises CLAP to comprehend the style, contents and emotions from a piece of audio. The main steps include:

\begin{enumerate}
    \item \textbf{Audio Segmentation}. Input song is split into multiple audio segments of varying lengths (Section \ref{sec:extraction}).
    \item \textbf{Audio Analysis}. Each segment is analysed by CLAP to extract segment-specific features, while the entire track is also analysed to determine its overall style and mood (Section \ref{sec:class_labels}).
    \item \textbf{Script Generation via LLM}. The extracted features are passed to a large language model (LLM) with a custom prompt instructing it to produce a music video script. The script comprises succinct scene descriptions (Section \ref{sec:script}).
    \item \textbf{Text to Video Generation}: Each scene description from the LLM is used to prompt a diffusion-based text-to-video model, generating video clips (Section \ref{sec:generating_videos}).
    \item \textbf{Final Assembly}. The generated video clips are concatenated and overlaid with the original audio, yielding the final music video.
\end{enumerate}
This multi-step strategy leverages the combined strengths of audio understanding (via CLAP), narrative construction (via LLM), and visual synthesis (via diffusion models) to produce a thematically consistent and music-driven video while leaving room for explainability, as the results of every intermediate step can be accessed and analysed.

\subsubsection{Identifying and Selecting Relevant Class Labels for CLAP}
\label{sec:class_labels}
Once the audio track is segmented, each segment is analysed with CLAP to obtain semantic class labels. We use three types of class labels:

\begin{itemize}
    \item \textbf{Segment-wise Class Labels}. These labels capture attributes relevant to each segment’s momentary content and are based on concepts from musical theory \cite{muller_computational_2016, cancino-chacon_computational_2018}. Some examples include: \textit{instrumental intensity}, \textit{prominent elements}, \textit{dynamic shifts}, and \textit{rhythmic and transitional functions}. By focusing on these localised attributes, segment-wise labels enable the system to highlight the specific musical nuances that define each clip.
    \item \textbf{Overall Content Style}. At the track level, we derive broader attributes such as \textit{genre}, \textit{tempo range}, and \textit{mood}. These attributes are chosen to help the LLM maintain a consistent theme or storyline across the entire video.
    \item \textbf{Overall Visual Style}. We also provide CLAP with prompts to generate or confirm a suitable overall visual style informed by the musical mood.
\end{itemize}
By combining segment-specific and track-wide analyses, we obtain a set of descriptors that guide the subsequent script generation and video synthesis steps.

\subsubsection{Music Video Script Generation}
\label{sec:script}
We deploy a reasoning-LLM to produce a structured music video script. This script details each scene’s visuals in a concise, thematically consistent manner, reflecting both the segment-wise attributes (e.g. changes in instrumentation or intensity) and the track-wise mood and style. The LLM prompt is constructed to include:

\begin{itemize}
    \item \textbf{Story or Structure Cues}. Encouraging the LLM to generate a coherent narrative.
    \item \textbf{Characters}. If desired, specify the type and number of characters (human or otherwise) to appear in the video.
    \item \textbf{Technical Constraints}. Number of scenes, we specifically instruct it to use a maximum of one sentence per scene description.
    \item \textbf{Audio-based Context}. Placeholders where the CLAP-derived labels are injected to inform the scene content.
    \item \textbf{Stylistic Guidelines}. Overall visual style (\textit{colour palette}, \textit{atmosphere}) that is consistent with the track’s mood.
\end{itemize}

An example of the prompt structure in pseudocode can be found in the Supplementary Material, together with examples of a prompt generated this way and its corresponding response. This approach ensures that the final output is straightforward to parse, limiting extraneous content that might confuse the text-to-video model in subsequent steps. Each scene description corresponds directly to one audio segment within the overall timeline. This way of parsing enables providing each scene prompt into the text-to-video model without further manual intervention or editing.

\begin{figure*}[ht]
\centering
\begin{subfigure}[b]{0.33\linewidth}
    \centering
    \includegraphics[width=\linewidth]{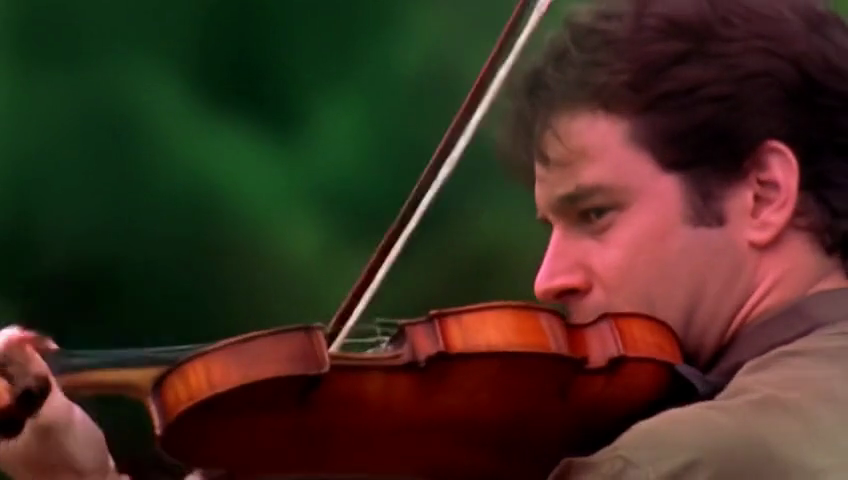}
    \caption{Classical}
\end{subfigure}
\hfill
\begin{subfigure}[b]{0.33\linewidth}
    \centering
    \includegraphics[width=\linewidth]{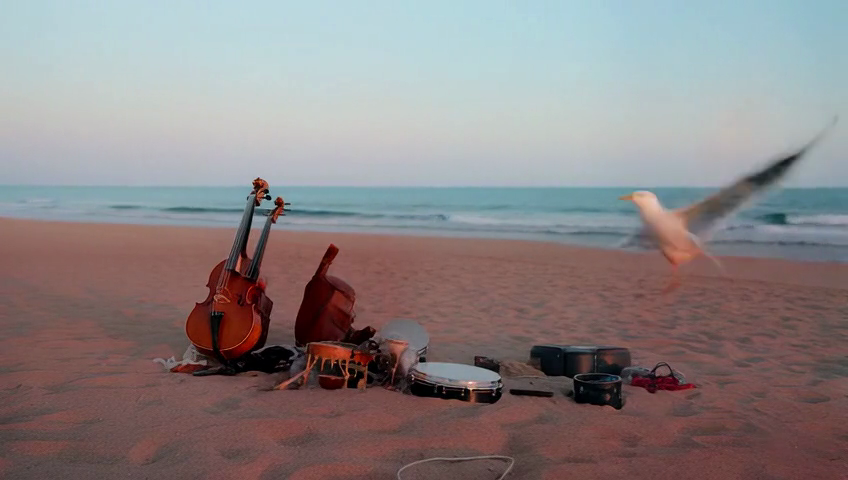}
    \caption{Jazz}
\end{subfigure}
\hfill
\begin{subfigure}[b]{0.33\linewidth}
    \centering
    \includegraphics[width=\linewidth]{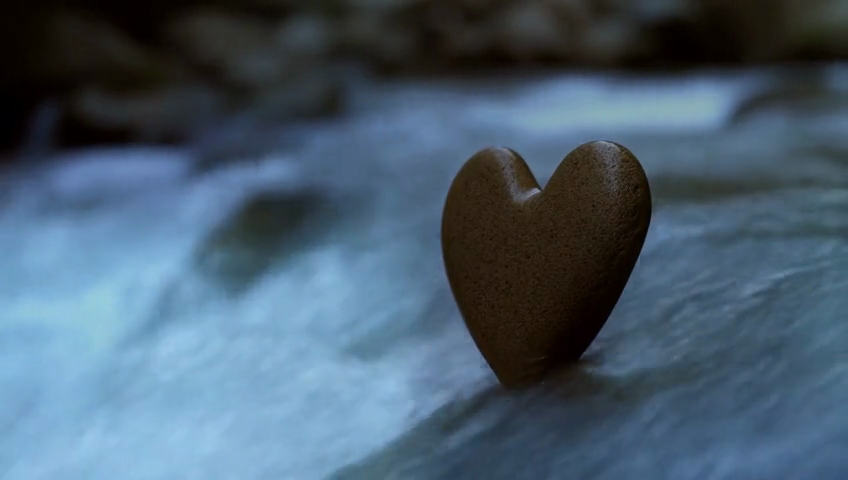}
    \caption{Pop}
\end{subfigure}

\vspace{1em} 

\begin{subfigure}[b]{0.33\linewidth}
    \centering
    \includegraphics[width=\linewidth]{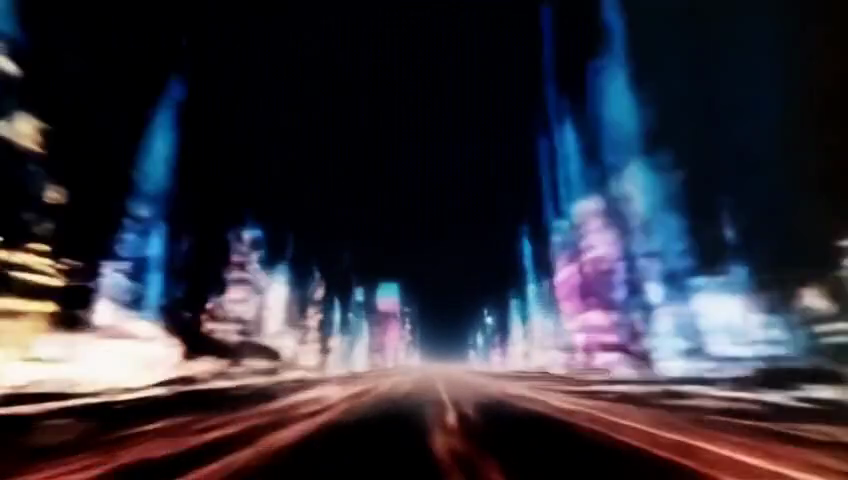}
    \caption{Heavy Metal}
\end{subfigure}
\hfill
\begin{subfigure}[b]{0.33\linewidth}
    \centering
    \includegraphics[width=\linewidth]{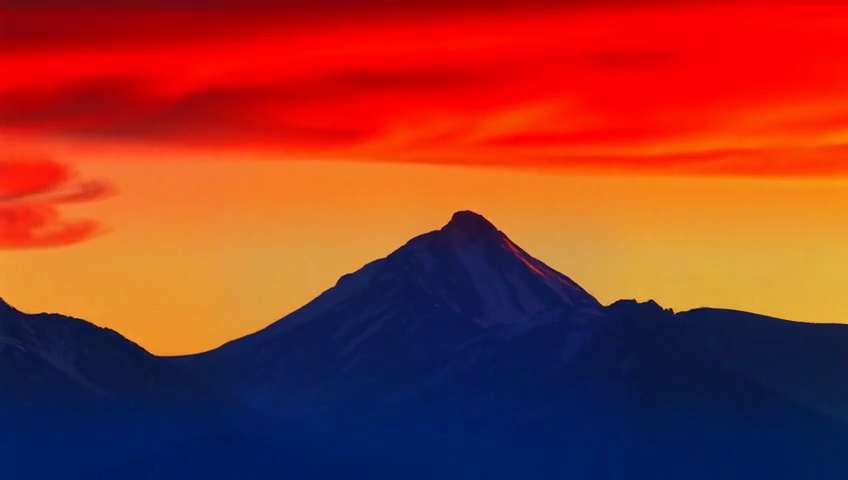}
    \caption{Old Blues}
\end{subfigure}
\hfill
\begin{subfigure}[b]{0.33\linewidth}
    \centering
    \includegraphics[width=\linewidth]{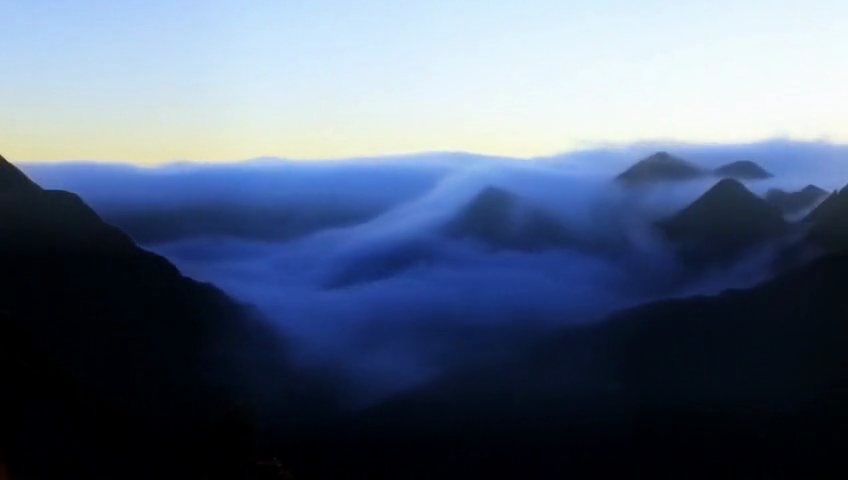}
    \caption{Vocal Jazz}
\end{subfigure}

\caption{Screenshots captured from final video outputs featuring the songs of specified genres. As hinted by the screenshots, the style and content of the results vary based on the subjectively perceived style and mood of the input song.}
\label{fig:commercial_music_comparison}
\end{figure*}
\subsection{Pipeline 2: LALM-based approach}
This alternative pipeline explores a more integrated approach to music video generation by leveraging the capabilities of an LALM. Unlike the CLAP-based method, which relies on explicit audio feature extraction and subsequent LLM prompting, this approach aims to derive the visual narrative directly from the LALM's ability to use a piece of audio as input. The LALM is prompted to generate a concise short story or narrative arc that thematically and emotionally aligns with the given song. This story serves as the conceptual blueprint for the entire music video, aiming to capture the song's essence in a coherent textual form.

The generated story is then used to inform the creation of individual scene descriptions. We achieve this by prompting the LALM itself to break down its story into segment-aligned visual cues, and then using a reasoning LLM to interpret the story and generate prompts for each pre-determined audio segment (as per the method in Section~\ref{sec:extraction}). The goal is to translate the overarching narrative into concrete scene descriptions that correspond to the song's temporal flow. Similar to the first approach, each scene description derived from the LALM's story is fed into a diffusion-based text-to-video model to generate corresponding video clips. These clips are then concatenated in chronological order and overlaid with the original audio track to produce the final music video.

This LALM-based approach offers the potential for more creatively coherent and emotionally resonant music videos, as the entire visual narrative stems from a single, integrated understanding of the audio. It reduces the need for manual feature engineering (like CLAP labels) and allows the model to infer relationships between audio elements and visual concepts without guidance, eliminating the bias introduced by overly specific human instructions. However, it relies heavily on the LALM's ability to accurately interpret musical nuances and translate them into compelling narratives, which can be less predictable than explicit feature-based prompting.

\subsection{Shared pipeline components}
\subsubsection{Segmentation}
\label{sec:extraction}
To segment the audio track, we experiment with two approaches. The first is a randomised segmentation method, producing segments of durations (randomly) varying between 4 and 8 seconds in length. Additionally, we developed a rule-based system which attempts to mimic how a human video editor might decide on cuts in a music video. The designed system makes every cut occur due to one of three possible factors:

\begin{itemize}
    \item Significant frequency changes in the music.
    \item The passage of a certain number of beats since the last segment change.
    \item If the current segment has reached a predefined maximum duration of 7 seconds, forcing a cut.
\end{itemize}

\subsubsection{Generating Videos}
\label{sec:generating_videos}
We generate video clips given the textual descriptions. For this task, we employ two diffusion-based text to video models, \textit{mochi-1} (CLAP-based pipeline) and \texttt{WAN 2.1} (LALM-based pipeline). Each scene description, usually limited to a single sentence, is the input prompt for the model. We try to mitigate this limitation by including overall style guidelines that help maintain a unified look. After each clip is generated, the results are concatenated following their original chronological order and overlaid with the music track.
\section{Experiments}
In this section, we present our experimental setup and evaluate the performance of the pipeline on a diverse set of musical examples. Pipelines were run using an NVIDIA H100 GPU with 80 GB of VRAM.
To evaluate the pipeline's robustness and generalizability, we assembled a diverse set of music tracks that cover a wide range of styles, tempos, and thematic contexts. This set includes self-collected, non-studio-mastered Irish folk music and a variety of commercially available tracks from genres such as `blues', `jazz', `heavy metal', and `classical music'. This diversity allows for a comprehensive assessment of how the system handles different audio characteristics and artistic intents. Figure \ref{fig:commercial_music_comparison} shows sample screenshots taken from various generated videos. For a more comprehensive understanding of our results, we invite the reader to visit our dedicated \href{https://github.com/goodPointP/Results-For-Music-Visualization-Generation-Pipeline}{Github repository}\footnote{https://github.com/goodPointP/Results-For-Music-Visualization-Generation-Pipeline} where a selection of generated music videos and the source code are publicly available. 

We conducted a within-subject evaluation with an ad-hoc sample consisting of five participants who were unaffiliated with the project to assess the narrative quality and visual coherence of videos generated by the two generative pipelines. The study was approved by the IT University of Copenhagen Ethics committee (No. 2023 – 1767-1217861).

\subsection{Exploratory Generative Pipeline Evaluation}
\label{sec:expl_eval}
Each participant watched a total of six music videos, with three generated by CLAP (genres: Jazz, Vocal Jazz, Traditional Irish Folk Song) and three by LALM (genres: Pop, Vocal Jazz, Heavy Metal). The videos were presented in randomised order, and the participants were not informed of which pipeline generated which video.

\begin{figure}[tb]
  \centering
  \includegraphics[width=\linewidth]{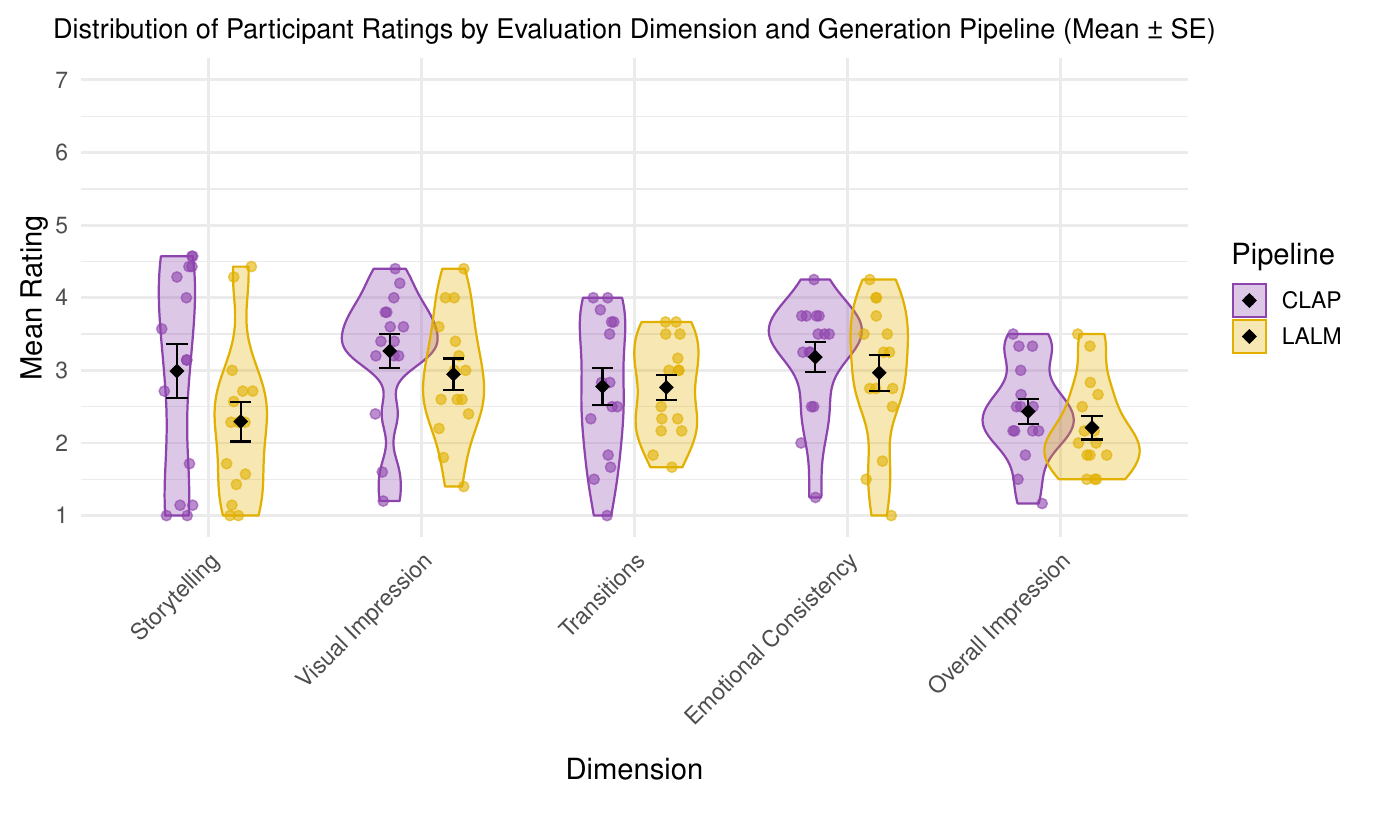}
  \caption{Participant Ratings by Dimension and Generation Pipeline.
  Violin plots show the distribution of participant ratings across five evaluation dimensions (\textit{Storytelling}, \textit{Visual Impression}, \textit{Transitions}, \textit{Emotional Consistency}, and \textit{Overall Impression}) for the Contrastive language–audio pretraining (CLAP) and Large Audio Language Model (LALM) video generation pipelines. Each dot represents a participant’s average rating for a specific genre within a given pipeline. Black diamonds indicate the pipeline mean, and black vertical bars represent the standard error (\textit{SE}) around the mean. Ratings for \textit{Storytelling} range from 1 (low) to 7 (high), while all other dimensions use a 1–5 Likert scale.
  }
  \label{fig:rating_violin_plot}
\end{figure}

After each music video, participants completed a survey assessing five dimensions\footnote{Items for the survey were created by the authors.} on a 7-point likert scale (Strongly Disagree - strongly agree): Storytelling (7 items), Visual Impression (5 items), Transitions (6 items), Emotional Consistency (4 items), and Overall Impression (6 items). After each dimension, an open text box was presented for the participants to explain their reasoning behind the ratings. For each video, the responses were averaged per dimension to obtain a single rating per participant, video, and dimension. This resulted in 15 ratings per pipeline for each dimension. For the distribution of these ratings across all videos, grouped by pipeline and evaluation dimension, see Figure \ref{fig:rating_violin_plot}. A complete list of survey items is provided in Section \ref{sec:survey_items}.

\begin{table}[tb]
\centering
\caption{Means (\textit{M}) and standard deviations (\textit{SD}) of participant ratings per evaluation dimension and pipeline. Note. Ratings were averaged across 5 participants and 3 videos per pipeline (15 ratings per dimension, 75 entries per pipeline overall rating). CLAP = Contrastive Language–Audio Pretraining; LALM = Large Audio Language Model.}
\label{tab:dimension_means}
\begin{tabular}{lccccc}
\toprule
Dimension & \multicolumn{2}{c}{CLAP} & \multicolumn{2}{c}{LALM} \\
 & \textit{M} & \textit{SD} & \textit{M} & \textit{SD} \\
\midrule
Emotional Consistency & 3.18 & 0.79 & 2.97 & 0.96 \\
Overall Impression    & 2.43 & 0.67 & 2.21 & 0.64 \\
Storytelling          & 2.99 & 1.43 & 2.30 & 1.06 \\
Transitions           & 2.78 & 0.98 & 2.77 & 0.67 \\
Visual Impression     & 3.27 & 0.90 & 2.95 & 0.84 \\
Overall Rating     & 2.93 & 1.01 & 2.64 & 0.89 \\
\bottomrule
\end{tabular}
\end{table}

On average, CLAP achieved a higher overall rating (\textit{M} = 2.93, \textit{SD} = 1.01) than LALM (\textit{M} = 2.64, \textit{SD} = 0.89). For the mean and standard deviation of participant ratings for each dimension and pipeline, see Table \ref{tab:dimension_means}. Visual inspection of the plot in Figure \ref{fig:rating_violin_plot} seems to indicate that the CLAP model tends to score slightly higher across all dimensions than the LALM pipeline. As the sample of five participants and six music videos is too small to draw meaningful conclusions, these results should be considered exploratory. Genre was not controlled across pipelines, so differences may also reflect subjective preferences and preconceptions about what videos in these genres should look like, rather than differences in generation quality. The sample size also prevents internal validity checks for the self-created rating dimensions, which should be addressed in future work, along with the open-ended responses.

\subsection{Qualitative Evaluation}
To evaluate the effectiveness of the generated music videos, a preliminary interview was conducted with two participants who reviewed all steps of the pipeline outputs. One of the participants interviewed is a non-expert, and one is a professional videographer. When reviewing the video outputs, the feedback revealed several key areas for improvement, primarily centred on the lack of visual and narrative consistency, supporting the low ratings for dimensions in the quantitative ratings. A recurring critique was that the videos lacked a unifying artistic vision, with inconsistent colour palettes, lighting, and overall visual styles from one shot to the next. This disjointedness was reported to feel like a collection of stock footage, which prevented viewers from becoming emotionally invested. This issue was compounded by a lack of character consistency, where protagonists would frequently change appearance or even species (\href{https://github.com/goodPointP/Results-For-Music-Visualization-Generation-Pipeline?tab=readme-ov-file#vocal-jazz-1}{"Vocal Jazz, LALM-Based method" example}\footnote{https://github.com/goodPointP/Results-For-Music-Visualization-Generation-Pipeline?tab=readme-ov-file\#vocal-jazz-1}) between shots, breaking the narrative thread.

When reviewing the text outputs, such as the narrative concepts and screen scripts, participants also offered feedback on the AI-generated stories that guided the video generation in the LALM-based pipeline. The high-level narrative concepts were generally well-received and regarded as fitting to the music in question. A more nuanced critique concerned the writing style; a uniform, matter-of-fact tone was perceived as effective for some genres (e.g., metal) but felt emotionally detached for more narrative-driven concepts. Participants suggested that these scripts would be more compelling with a 'literary flair' tailored to the specific mood of the music.

A key recommendation was to provide the video model with stronger and more persistent stylistic prompts, such as a predefined colour palette or specific aesthetic references, to ensure a unified look and feel. Furthermore, feedback indicated the need for mechanisms to maintain character identity across all generated clips and to adapt the narrative style of the script to the song's genre. Despite these critiques, participants noted that the AI was often successful at capturing the general mood and theme of the music without access to the lyrics, suggesting that the core concept of the pipeline is promising but requires further refinement in execution to achieve narrative and visual cohesion.
\section{Conclusions}

\subsection{Summary of Findings}

Our results show that both CLAP and LALMs can effectively extract meaningful audio features from various musical inputs, and when combined with a reasoning-focused LLM and concise prompts, yield coherent, stylistically aligned video scripts. This is based on the fact that in the interviews, the narrative concepts were perceived as fitting the music, unlike the resulting video outputs. Although the diffusion model can effectively visualise these scripts, it often lacks visual consistency. Overall, these findings highlight the potential of integrating latent audio feature extraction techniques with LLM-driven text to video generation for creating compelling, conceptually coherent music videos.

\subsection{Limitations}

When the generated scripts involve recurring characters (especially human characters), the text-to-video model often struggles to maintain visual consistency across scenes. Every new clip often differs from the one before in style, motion and colour. With characters, the variations in facial features, clothing, or style can undermine narrative continuity.



\subsection{Future Work}

Looking ahead, we identify several research directions that could significantly enhance the quality, responsiveness, and ethical considerations of our pipeline:

\textbf{Improving visual consistency.} As reported by the survey participants and the conducted interview, improving character consistency would have a positive impact on storytelling and thus the coherency of the final video.

\textbf{Incorporating lyrics.} Incorporating lyrics into the prompt could deepen the LLM’s understanding of the track’s narrative and might lead to scene descriptions that more accurately reflect lyrical themes or storylines.





\textbf{AI and Art: Creative Collaboration and Authorship.} Exploring how human creators can guide AI’s creativity, and how AI’s role may evolve in shaping artistic expression.

\textbf{Condoning a larger, expanded user study.} To better understand where differences in AI-generated video quality arise, future evaluations could assess intermediate outputs of the generation process. This would help clarify whether using specialised models for specific steps, such as emotion detection, script generation, or video synthesis, results in higher-quality outputs or whether general-purpose models can handle the full pipeline. It would also be useful to conduct more in-depth interviews to explore how people interpret and evaluate AI-generated video content. This is especially relevant, as prior work has shown that perceptions of AI are shaped not just by task performance, but also by underlying expectations \cite{kosch_placebo_2022, kloft_ai_2024}, preconceptions \cite{cave_scary_2019}, transparency \cite{liu2021TrustAI}, and narratives \cite{hermann2020beware}. Future work could also include comparisons to professionally produced or human-curated videos to better understand how generated content is judged relative to real-world storytelling standards.

By addressing these directions, we aim to develop an approach that is aesthetically compelling and contextually sensitive, ultimately empowering artists, content creators, and end-users to generate music videos that are aligned with their musical inspiration.


\section*{Acknowledgement}
This work is funded by the European Union within the Horizon Europe research and innovation programme under grant agreement No. 101136006 – XTREME project, coordinated by S.S. Brandt/IT University of Copenhagen, Denmark. Views and opinions expressed are however those of the authors only and do not necessarily reflect those of the European Union, European Union can not be held responsible for them. This project was partially funded by the Pioneer Centre for AI, DNRF grant number P1.
{
    \small
    \bibliographystyle{ieeenat_fullname}
    \bibliography{main,references_zotero_Xtreme_AVC,JSreferences}
}

\clearpage
\setcounter{page}{1}
\twocolumn[{%
\renewcommand\twocolumn[1][]{#1}%
\maketitlesupplementary

}]
\section{Prompt Flow}
\label{sec:prompt}
\subsection{CLAP-based Approach}
In the following, we present the prompt structure for the CLAP-based approach in pseudocode in Block \ref{block:prompt_structure}, an example prompt in Block \ref{block:prompt}, and the corresponding response in Block \ref{block:response}.
\vspace{0pt} 

\begin{minipage}{\textwidth}%
\begin{lstlisting}[caption=Prompt structure for the LLM used in the CLAP-based approach, label=block:prompt_structure, style=mystyle]
number_of_scenes = length(segments)
    
final_prompt = "You need to think of a story and a script for a music video. {additional_prompt}
                The story structure needs to have a beginning, middle and an ending.
                You will write the story based on the characteristics of the music that are provided to you.
                The story needs to be told as descriptions of sccenes that will appear in a music video.
                The structure needs to be reflected in the scene layout, to try to mimic the story's progression.
                The story should include at least one animal character. For consistency, repeat the animal's description in every scene.
                Make sure the scene descriptions are concrete and to the point, they need to be easy for a text2video model to generate videos from.
                There are {number_of_scenes} scenes in total.
                When listening to the entire song, it can be described like this:
                the overall {overall_song_description_keywords}
                When listening to individual segments, they can be described like this:
                {scene_descriptions}
                Each scene needs to visually match the following guidelines:
                {overall_song_description_keywords_visual_guides}
                Scene descriptions need to be as brief as possible. Every scene can be described in one sentence at the most. Omit all unnecesary details from the description.
                In your response, start the description of every scene with the exact letters: "SCENE #:", '#' substituted with the scene number. Do NOT add any special or any other type of characters to this line! Example: "SCENE 1:\\n"
                Before the start of the script, add the words "BEGIN SCRIPT\\n" so that I can easily extract it."
\end{lstlisting}%
\end{minipage}

\vspace{\fill}
~\\
\clearpage


\begin{minipage}{\textwidth}%
\begin{lstlisting}[caption=Prompt sent to the LLM based on CLAP features, label=block:prompt, style=mystyle]
You need to think of a story and a script for a music video.
The story structure needs to have a beginning, middle and an ending.
You will write the story based on the characteristics of the music that are provided to you.
The story needs to be told as descriptions of sccenes that will appear in a music video.
The structure needs to be reflected in the scene layout, to try to mimic the story's progression.
The story should include at least four people. For consistency, repeat their descriptions in every scene.
Make sure the scene descriptions are concrete and to the point, they need to be easy for a text2video model to generate videos from.
There are 7 scenes in total.
When listening to the entire song, it can be described like this:
the overall Instrumental energy is It has multiple peaks and valleys throughout.. Instrumental palette is Orchestral or cinematic instruments. Tempo range is Very fast (140+ BPM). Production quality is Very polished, glossy, and modern. Mood is Uplifting and bright. 

When listening to individual segments, they can be described like this:

Scene 1:
Instrumental intensity is Moderate intensity with a clear beat. Prominent element is String or orchestral elements. Dynamic shift is It fluctuates multiple times within the segment. Rhythm is Irregular or changing time signatures. Function in transitioning is It acts as a noticeable break or "breather". The scene will be 5.49 seconds long.
Scene 2:
Instrumental intensity is Moderate intensity with a clear beat. Prominent element is String or orchestral elements. Dynamic shift is It fluctuates multiple times within the segment. Rhythm is Irregular or changing time signatures. Function in transitioning is It features a sudden drop or pause before the next section. The scene will be 7.13 seconds long.
Scene 3:
Instrumental intensity is High energy and full instrumentation. Prominent element is String or orchestral elements. Dynamic shift is It fluctuates multiple times within the segment. Rhythm is Irregular or changing time signatures. Function in transitioning is It features a sudden drop or pause before the next section. The scene will be 7.87 seconds long.
Scene 4:
Instrumental intensity is High energy and full instrumentation. Prominent element is String or orchestral elements. Dynamic shift is It fluctuates multiple times within the segment. Rhythm is Irregular or changing time signatures. Function in transitioning is It cleanly continues the energy from the previous segment. The scene will be 6.66 seconds long.
Scene 5:
Instrumental intensity is Moderate intensity with a clear beat. Prominent element is String or orchestral elements. Dynamic shift is It stays uniformly loud/energetic. Rhythm is Irregular or changing time signatures. Function in transitioning is It slowly fades out or prepares for a drop. The scene will be 6.2 seconds long.
Scene 6:
Instrumental intensity is High energy and full instrumentation. Prominent element is Synths or electronic sounds. Dynamic shift is It fluctuates multiple times within the segment. Rhythm is Irregular or changing time signatures. Function in transitioning is It dramatically shifts the energy or mood. The scene will be 5.72 seconds long.
Scene 7:
Instrumental intensity is High energy and full instrumentation. Prominent element is String or orchestral elements. Dynamic shift is It fluctuates multiple times within the segment. Rhythm is Irregular or changing time signatures. Function in transitioning is It features a sudden drop or pause before the next section. The scene will be 4.94 seconds long.

Each scene needs to visually match the following guidelines:
Location is exterior. Visual setting is Natural. Visual style is Monochromatic or limited color palette. Visual focus is Multiple focal points or characters. 

Scene descriptions need to be as brief as possible. Every scene can be described in one sentence at the most. Omit all unnecesary details from the description.
In your response, start the description of every scene with the exact letters: "SCENE #:", '#' substituted with the scene number. Do NOT add any special or any other type of characters to this line! Example: "SCENE 1:\n"
Before the start of the script, add the words "BEGIN SCRIPT\n" so that I can easily extract it.
\end{lstlisting}%
\end{minipage}
~
\qquad 
~\\
\newpage

\begin{minipage}{\textwidth}%
\begin{lstlisting}[caption=Prompt response from LLM based on CLAP features, label=block:response, style=mystyle]
Okay, so I'm trying to help this user who wants a story and a script for a music video. Let me break down what they need. They provided a lot of specific details about the song's structure, the scenes, and the visual guidelines. 

First, the story needs to have a beginning, middle, and ending. It should be told through scene descriptions, and each scene needs to reflect the song's segments. They mentioned that there are seven scenes, each with specific characteristics like instrumental intensity, prominent elements, dynamic shifts, and transition functions. Plus, each scene should be a single sentence, starting with "SCENE #:".

The user also provided guidelines for the visual style: exterior locations, natural settings, monochromatic or limited color palette, and multiple focal points or characters. The story should include at least four people, with consistent descriptions across scenes. The production quality is polished and modern, with an uplifting and bright mood.

I need to make sure each scene's description is concrete and easy for a text-to-video model to generate. The scenes should transition smoothly, reflecting the dynamic shifts and energy changes of the song. Also, the locations should be natural, like forests, mountains, or beaches, to fit the exterior requirement.

I'll start by outlining each scene based on the provided structure, ensuring that each one matches the corresponding segment's characteristics. I'll include four main characters, each with distinct roles and appearances, to keep the story consistent. The descriptions should be brief, focusing on the main visual elements without unnecessary details.

Finally, I'll structure the script with "BEGIN SCRIPT" at the start and each scene clearly labeled. I'll make sure the transitions between scenes follow the energy and mood shifts indicated by the song's segments. This should result in a cohesive and visually engaging music video script that aligns perfectly with the provided song details.
</think>

BEGIN SCRIPT  

SCENE 1: A group of four people, dressed in vibrant, flowing outfits, stand at the edge of a vast, sun-drenched forest. Strings of light shimmer in the distance as they begin to walk forward, their movements synchronized to the rhythm.  

SCENE 2: The group reaches a clearing where a lone figure, dressed in a shimmering silver cloak, stands. The figure raises their arms, and the strings of light twist into a swirling vortex above. The group steps into the light, their colors blending together as the strings pull them toward the sky.  

SCENE 3: The group emerges from the light into a futuristic, neon-lit cityscape. Skyscrapers glow with shifting colors, and a massive screen displays abstract patterns. The lone figure from before now stands at the center, their cloak glowing faintly. The group surrounds them, their movements now more chaotic and energetic.  

SCENE 4: The city transforms around them. Buildings collapse into sand, and the sky darkens slightly. The lone figure raises their hands, and the sand reforms into a beach with crashing waves. The group stands amidst the waves, their expressions reflective and determined.  

SCENE 5: A single figure, now isolated on the beach, begins to fade into the horizon. The group watches, their forms becoming more abstract as the scene transitions. The last shot lingers on the group, their forms dissolving into light as the music fades.  

SCENE 6: The group reappears in a serene mountain valley. Snow-covered peaks rise into the distance, and a soft glow surrounds them. The lone figure, now wearing a different outfit, leads the group toward a crystal cave at the base of the mountains.  

SCENE 7: The group enters the cave, the walls shimmering with faint, ethereal light. The lone figure raises their hands, and the cave fills with a blinding light. The group steps forward, their forms dissolving into light as the music crescendos to a final, uplifting note.  

END SCRIPT
\end{lstlisting}%
\end{minipage}


\newpage
\onecolumn

\subsection{LALM-based Approach}
Block \ref{block:prompt_lalm} presents a prompt used when inquiring a LALM model to write a video script. Block \ref{block:response_lalm} is an example of a final video script generated by the LALM.

\begin{minipage}{\textwidth}%
\begin{lstlisting}[caption=Analysis prompt for the LALM model, label=block:prompt_lalm, style=mystyle]
lalm_conversation = [
        {'role': 'system', 'content': 'You are a helpful assistant.'}, 
        {"role": "user", "content": [
            {"type": "audio", "audio_url": song_audio},
            {"type": "text", "text": f"Think of a story for a music video for this song. {additional_user_prompt if additional_user_prompt else ''} The story should be engaging and visually appealing, with a focus on the emotions conveyed by the music and lyrics."},
        ]},
    ]
\end{lstlisting}%
\end{minipage}

\begin{minipage}{\textwidth}%
\begin{lstlisting}[caption=LALM-generated video script, label=block:response_lalm, style=mystyle]
In the music video the protagonist is a young woman named Elina who dreams of becoming a professional dancer. She faces numerous obstacles and rejections but remains determined to pursue her passion. One day, while practicing her routine alone in an empty studio, she hears the track playing in the background. The music Empowers her, and she feels a surge of confidence. As the bass drops, she begins to dance wildly, letting go of all her worries. The music's message of perseverance and self-empowerment resonates deeply with her, and she knows that she has found her calling. The video captures her journey from despair to triumph, showcasing her growth and resilience through the power of dance and music.
\end{lstlisting}%
\end{minipage}

\newpage
\onecolumn

\section{Exploratory Generative Pipeline Evaluation: Survey Items}
\label{sec:survey_items}

In the following, we present the survey items used for the preliminary evaluation described in \autoref{sec:expl_eval}. All items were rated on a 7-point Likert scale: where 1 = Strongly Disagree, and 7 = Strongly Agree.

\subsection*{Storytelling}
\begin{tabularx}{\textwidth}{@{}lX@{}}
\textbf{1.} & The video featured consistent characters. \\
\textbf{2.} & I felt connected to at least one of the characters. \\
\textbf{3.} & The narrative had a clear beginning, middle, and end. \\
\textbf{4.} & The characters had clear goals or challenges. \\
\textbf{5.} & The progression of the story felt motivated by the music. \\
\textbf{6.} & The characters’ actions were easy to follow. \\
\textbf{7.} & The storyline was easy to understand. \\
\end{tabularx}

\vspace{1em}
\textbf{Open Text Box:}  
\emph{What was your impression of the story and character development? What stood out to you—whether positive or negative?}

\vspace{1.5em}
\subsection*{Visual Impression}
\begin{tabularx}{\textwidth}{@{}lX@{}}
\textbf{1.} & The visual style of the video was aesthetically appealing. \\
\textbf{2.} & The visual style of the video was distracting or confusing. \textit{(reverse-coded)} \\
\textbf{3.} & The visual elements supported the mood of the music. \\
\textbf{4.} & The scene changes in the video matched the rhythm and structure of the music. \\
\textbf{5.} & The visual setting and mood were appropriate for the music. \\
\end{tabularx}

\vspace{1em}
\textbf{Open Text Box:}  
\emph{What was your impression of the visual style of the video? What stood out to you—whether positive or negative?}

\vspace{1.5em}
\subsection*{Video Cutting / Transitions}
\begin{tabularx}{\textwidth}{@{}lX@{}}
\textbf{1.} & Transitions between scenes felt natural. \\
\textbf{2.} & I noticed cuts that felt abrupt or distracting. \textit{(reverse-coded)} \\
\textbf{3.} & The timing of scene changes matched the rhythm of the music. \\
\textbf{4.} & The timing of scene changes matched the narrative flow. \\
\textbf{5.} & The pacing of the video supported the unfolding of the story. \\
\textbf{6.} & Some scene changes felt disconnected. \textit{(reverse-coded)} \\
\end{tabularx}

\vspace{1em}
\textbf{Open Text Box:}  
\emph{What was your impression of the pacing and transitions between scenes? What stood out to you—whether positive or negative?}

\vspace{1.5em}
\subsection*{Emotional Consistency}
\begin{tabularx}{\textwidth}{@{}lX@{}}
\textbf{1.} & The emotions conveyed in the video matched the emotional tone of the music. \\
\textbf{2.} & The emotional tone of the video changed abruptly or felt disjointed. \textit{(reverse-coded)} \\
\textbf{3.} & I felt emotionally engaged with the video throughout. \\
\textbf{4.} & The emotional tone remained consistent across different scenes. \\
\end{tabularx}

\vspace{1em}
\textbf{Open Text Box:}  
\emph{What was your impression of the emotional consistency of the video? What stood out to you—whether positive or negative?}

\vspace{1.5em}
\subsection*{Overall Video Impression}
\begin{tabularx}{\textwidth}{@{}lX@{}}
\textbf{1.} & I enjoyed the video. \\
\textbf{2.} & The video matched the emotional tone of the music. \\
\textbf{3.} & The video felt repetitive or boring at times. \textit{(reverse-coded)} \\
\textbf{4.} & I found the video creatively engaging. \\
\textbf{5.} & I would share or recommend this video. \\
\textbf{6.} & I took the video seriously as a creative work. \\
\end{tabularx}

\vspace{1em}
\textbf{Open Text Box:}  
\emph{Please reflect on the video’s overall quality. What aspects worked well? Where did it fall short or feel inconsistent with the music or prior descriptions?}

\twocolumn

\end{document}